\documentclass[aps, prd, twocolumn,showpacs, superscriptaddress, groupedaddress]{revtex4-2} 
\usepackage{graphicx}
\usepackage{amsmath}
\usepackage{amssymb}
\usepackage{dcolumn}
\usepackage{color}
\usepackage{xcolor}
\usepackage{placeins}
\usepackage{subfigure, rotating, bm, array}
\usepackage[pagebackref=false, colorlinks=true]{hyperref}
\usepackage{parskip}
\usepackage{ragged2e}
\hypersetup{
linkcolor=blue,     % color of internal links
citecolor=blue,     % color ofthe links to bibliography
urlcolor=red}      % color of  url
\begin{document}
\title{Thermodynamic Phase Transitions and Quantum Entropy Corrections in the Simpson–Visser Regular Black Hole}
\author{Vinayak Joshi}
\email{vinayak\_j@ph.iitr.ac.in}
\affiliation{Department of Physics, Indian Institute of Technology Roorkee, Roorkee 247667, Uttarakhand, India}
\author{Ashok B. Joshi}
\email{gen.rel.joshi@gmail.com}
\affiliation{International Centre for Space and Cosmology, School of Arts and Sciences, Ahmedabad University, Ahmedabad 380009 (Gujarat), India}
\affiliation{PDPIAS, Charotar University of Science and Technology, Anand 388421 (Gujarat), India}

\begin{abstract}
Regular black holes offer a compelling framework to explore the consequences of resolving the central singularity of standard black holes. Using the Simpson–Visser ``black-bounce" geometry as an elegant, analytically tractable framework, we explore the intricate thermodynamic behavior in such models. We demonstrate that this regular spacetime exhibits a critical instability, marked by a phase transition where the heat capacity is discontinuous. This transition signals a fundamental change in the black hole's evaporation state, which depends on the regularization parameter. Pushing beyond the semiclassical limit, we then derive the leading-order quantum corrections to the entropy via the Hamilton-Jacobi tunneling formalism. Our analysis provides a refined statistical basis for the entropy of non-singular spacetimes and offers a quantitative analysis of the nature of the black hole end-state. These results reveal that singularity resolution is not merely a geometric modification but a profound thermodynamic event, with direct implications for the stability and ultimate fate of evaporating black holes.

$\boldsymbol{key words}$ : Black hole Thermodynamics, Quantum Gravity
\end{abstract}
\maketitle

\section{Introduction}
The unresolved tension between general relativity and quantum mechanics stands as the central challenge of theoretical physics. While general relativity provides a description of gravity on macroscopic scales \cite{Misner:1973prb}, its prediction of spacetime singularities within black holes and at the cosmological origin marks the definitive boundary of its domain \cite{Penrose:1964wq, Hawking:1970zqf}. The occurrence of infinite curvature indicates a breakdown of classical general relativity in the high-energy regime, making the inclusion of quantum effects essential for a consistent description. Understanding how such quantum corrections resolve the singularity is therefore a central task, and black holes provide a natural setting for investigating these issues. Within this context, the semiclassical approximation has yielded profound insights, most notably the discovery of black hole thermodynamics \cite{Bekenstein:1973ur,Hawking:1975vcx}.

The concepts of Bekenstein-Hawking entropy and Hawking temperature has shown a remarkable link between horizon area, surface gravity, and the laws of thermodynamics, suggesting a statistical-mechanical origin for spacetime geometry. Recent developments have further strengthened this perspective, highlighting the continuing relevance of black hole thermodynamics to the understanding of spacetime and gravitational dynamics \cite{Strominger:1996sh,Padmanabhan:2009vy,Ashtekar:1997yu,Jacobson:2015hqa}. However, this paradigm relies on a singular background while simultaneously predicting the black hole's complete evaporation, a process that would inevitably involve the singular region and exacerbate the information loss paradox \cite{Giddings:1995gd,Unruh:2017uaw}. Therefore, a framework where the singularity is resolved from the beginning is necessary for a self-consistent description of the black hole life cycle. This necessity has given rise to the study of regular black holes, which serve as phenomenological models to explore the physical consequences of singularity resolution. 

An additional motivation comes from the fate of inner horizons. Strong cosmic censorship links physical predictability to the instability of Cauchy horizons, and recent analyses show that even in regular or quantum-corrected black holes these inner horizons typically undergo exponential mass amplification under perturbations \cite{Cardoso:2017soq,Bonanno:2020fgp,Carballo-Rubio:2021bpr,Bonanno:2022rvo,Bambi:2023try}. Moreover, mass-inflation–type instabilities can arise even in dynamical spacetimes without a strict Cauchy horizon, provided an inner trapping horizon evolves slowly \cite{Carballo-Rubio:2024dca}. These results reinforce the view that long-lived, nonsingular black-hole spacetimes require a controlled modification of the interior structure. These effective geometries replace the central singularity with a regular, non-singular core \cite{Bardeen68,Hayward:2005gi}. Recent studies have investigated the dynamical formation of regular black holes including thin-shell and Oppenheimer–Snyder collapse as well as fully dynamical scenarios \cite{Bueno:2024dgm,Bueno:2024eig,Bueno:2025gjg}.  This regularity involves the introduction of matter that violates the Null Energy Condition (NEC). \cite{Visser:1995cc,Visser:1989kh,Morris:1988tu}. Regular rotating black holes have also been investigated as part of this broader effort to construct nonsingular spacetimes \cite{Bambi:2013ufa,Azreg-Ainou:2014pra,Torres:2022twv}. In particular, the work of Franzin et al. has demonstrated certain rotating regular black hole solutions whose inner horizon has zero surface gravity and is, therefore, stable against mass inflation \cite{Franzin:2022wai, Ghosh:2022gka}.
Distinct from geometric deformations, alternative regularization schemes have been developed that rely on a metric signature change across the event horizon. In this Lorentzian-Euclidean framework, the transition to an imaginary time coordinate effectively prevents causal geodesics from extending to the central singularity. This dynamical mechanism, termed atemporality, ensures the geodesic completeness of the spacetime and preserves fundamental conservation laws \cite{Capozziello:2024ucm,DeBianchi:2025bgn,Capozziello:2025wwl}. Among other wide proposals of regular spacetime models \cite{Frolov:2016pav,Frolov:2017dwy,Bardeen:2018frm,Carballo-Rubio:2018pmi,Cano:2018aod}, the Simpson-Visser black bounce model \cite{Simpson:2018tsi,Lobo:2020ffi} offers a particularly compelling framework due to its elegant and minimalist construction. Governed by a single regularization parameter $a$, which introduces a characteristic length scale, it provides an analytic, continuous interpolation between the Schwarzschild black hole $(a\rightarrow0)$, a regular one-way wormhole known as a ``black-bounce," and a two-way traversable wormhole.

The black-bounce case is particularly interesting, as it represents an object that is asymptotically indistinguishable from a classical black hole yet possesses a fundamentally different internal structure, replacing the singularity with a smooth bounce to another asymptotic region.
The central idea motivating our present work is that a geometric alteration as fundamental as singularity resolution must manifest as a distinct thermodynamic signature. Its modified structure must be reflected in its stability and response to quantum fluctuations. The Simpson–Visser ‘black-bounce’ metric forms a broader and qualitatively different class of regular black holes, interpolating smoothly between the Schwarzschild solution and a Morris–Thorne wormhole. As emphasized in the original work by Simpson and Visser \cite{Simpson:2018tsi}, this geometry generalizes and broadens the space of regular black hole solutions beyond the models usually considered. This difference in geometric structure leads to distinct thermodynamic behaviour, including the emergence of a Davies-type phase transition and a modified entropy structure, which motivates a dedicated analysis separate from previous studies. While thermodynamic studies of regular black holes have largely focused on Bardeen and Hayward models \cite{Tharanath:2014naa,Maluf:2018lyu,Wu:2024gqi,PhysRevD.99.024015}, and recent work has examined the phase structure of Simpson--Visser black holes in AdS spacetime \cite{Kumar:2025nio}, two key questions remain unaddressed: whether the asymptotically flat Simpson--Visser geometry exhibits an intrinsic thermodynamic phase transition driven purely by singularity resolution, without recourse to a cosmological constant or extended phase space, and what form the quantum corrections to its entropy take. Our work answers both. We identify a Davies-type phase transition at $a_{\text{crit}} = \sqrt{2}\,m$ that is intrinsic to the regularized geometry itself, and we derive the quantum-corrected entropy of the Simpson--Visser black hole via the Hamilton-Jacobi tunneling formalism, revealing that the regularization parameter $a$ enters the entropy at leading order and simultaneously regulates the convergence of the perturbative quantum expansion.

Our investigation therefore proceeds along these two complementary lines of inquiry. First, we conduct the analysis of the black hole's thermodynamic stability. We go beyond the calculation of temperature and entropy to probe the system for critical behavior by examining the heat capacity and free energy as functions of the regularization parameter $a$. Our analysis reveals the existence of a critical point where the heat capacity diverges, signaling a thermodynamic stability transition marked by a Davies point. This result provides concrete evidence that the regularization scale acts as a control parameter that can drive the system between phases of stability and instability. Such phase transitions where the heat capacity tends to change sign at a turning point are also in other black hole models \cite{Jing:2008an,Croney:2025cue}. 
Second, we advance beyond the semiclassical area law to quantify the leading-order quantum corrections to the black hole entropy. While the Bekenstein-Hawking entropy is a semiclassical result, the logarithmic corrections, which encode information about the quantum state counting and fluctuations of the background geometry itself, are a crucial part of the full quantum picture. Employing the well-established Hamilton-Jacobi tunneling method \cite{Parikh:1999mf,Parikh:2004ih,Banerjee:2008cf,Majhi:2009pr}, we explicitly derive these corrections for the Simpson-Visser spacetime. This calculation provides a refined statistical basis for the entropy of a regular black hole and serves as a crucial step towards understanding the quantum structure of non-singular spacetimes.
Taken together, this dual analysis of classical thermodynamic stability and leading-order quantum corrections provides a complete thermodynamic portrait of a regular black hole. It demonstrates quantitatively how the removal of the singularity governs the object's phase structure and modifies its quantum entropy, offering critical insights into the nature of black hole remnants and the resolution of the information paradox.

The paper is structured as follows. In Section II, we review the Simpson-Visser geometry and derive its fundamental semiclassical thermodynamic quantities. Section III is dedicated to the detailed analysis of thermodynamic stability, culminating in the identification of the phase transition. In Section IV, we present the derivation of the quantum-corrected entropy via the tunneling formalism. We conclude in Section V with a synthesis of our results and a discussion of their broader implications for quantum gravity phenomenology. We adopt geometric units where 
$G=c=\hbar=k_B=1$.
\section{The Simpson-Visser Spacetime and its Thermodynamic Foundation}
\label{sec:geometry}

In this section, we establish the theoretical groundwork for our analysis. We begin by reviewing the essential geometric properties of the static Simpson--Visser spacetime, with particular attention to its regularity and causal structure. We then revisit the stress-energy tensor required to source this geometry and scrutinize its compliance with the standard energy conditions, providing physical context for its exotic nature. Finally, we compute the foundational semiclassical thermodynamic properties, the Hawking temperature and Bekenstein--Hawking entropy, which will serve as the baseline for the more detailed investigations of stability and quantum corrections in the subsequent sections. This section is based on the earlier works by Simpson and Visser \cite{Simpson:2018tsi, Simpson:2019cer}.

\subsection{Geometric Structure and Regularity}

The starting point of our investigation is the static, spherically symmetric line element proposed by Simpson and Visser \cite{Simpson:2018tsi}, which provides a minimalist and analytically tractable framework for modeling singularity resolution:
\begin{equation}
ds^2 = -f(r) dt^2 + \frac{dr^2}{f(r)} + \mathcal{R}(r)^2 (d\theta^2 + \sin^2\theta d\phi^2),
\label{eq:metric}
\end{equation}
where the metric function is given by
\begin{equation}
f(r) = 1 - \frac{2m}{\sqrt{r^2+a^2}},
\label{eq:metric_func}
\end{equation}
and the area radius is defined as $\mathcal{R}(r) = \sqrt{r^2+a^2}$. Here, $m$ represents the ADM mass of the spacetime, while the parameter $a \geq 0$ introduces a fundamental length scale responsible for the regularization. A key feature of this geometry is that the area of a 2-sphere, $A(r) = 4\pi \mathcal{R}(r)^2 = 4\pi(r^2+a^2)$, is bounded from below by a minimum value $A_{\text{min}} = 4\pi a^2$ at the ``bounce point'' $r=0$. This prevents the formation of a point-like singularity. It is essential to distinguish the role of $r=0$ in this geometry from that in the Schwarzschild spacetime. In the Schwarzschild solution, the radial coordinate is restricted to $r \in [0, \infty)$, and $r=0$ is the boundary of the manifold where the area of the 2-sphere vanishes and curvature invariants diverge, this is the curvature singularity. In the Simpson--Visser geometry, by contrast, the coordinate $r$ ranges over, $r \in (-\infty, +\infty)$. The point $r=0$ is not a boundary but a regular interior point of the manifold. The surface $r=0$ is therefore not a singularity but a smooth, finite-area ``bounce'' surface through which the geometry continues to negative $r$ values, corresponding to a separate asymptotic region.
The definitive test of a geometry's regularity is the behavior of its curvature invariants. For the metric \eqref{eq:metric}, the polynomial scalar invariants are finite provided $a > 0$. The Ricci scalar is calculated to be:
\begin{equation}
R = \frac{2a^2(3m - \sqrt{r^2+a^2})}{(r^2+a^2)^{5/2}}.
\label{eq:ricci_scalar}
\end{equation}
More comprehensively, the Kretschmann scalar $K = R_{\mu\nu\rho\sigma}R^{\mu\nu\rho\sigma}$ is also finite everywhere. As $r \to 0$, both scalars approach finite values, $R(0) = (6m-2a)/a^4$ and $K(0) = 12m^2/a^6$, explicitly demonstrating the absence of a curvature singularity at the spacetime's core  \cite{Simpson:2018tsi}. In the asymptotic limit $r \to \infty$, the spacetime is asymptotically flat, recovering the Schwarzschild behavior.

The causal structure is governed by the locations of horizons, determined by the roots of $f(r)=0$. This condition yields $\sqrt{r^2+a^2}=2m$, with solutions for the horizon radii given by $r_h = \pm\sqrt{4m^2-a^2}$. The physical nature of the spacetime is thus partitioned into four distinct classes based on the ratio $a/m$ \cite{Simpson:2018tsi}:
\begin{itemize}
    \item \textbf{Regular Black Hole ($0 < a < 2m$):} This is the case of interest for our thermodynamic analysis. The horizon condition $f(r) = 0$ requires $\mathcal{R}(r_+) = \sqrt{r_+^2 + a^2} = 2m$, yielding a coordinate location $r_+ = \sqrt{4m^2 - a^2}$. While the areal radius at the horizon is $\mathcal{R}(r_+) = 2m$ regardless of $a$, the coordinate position $r_+$ decreases with increasing $a$ and is strictly less than $2m$ for any $a > 0$. The bounce surface at $r = 0$ lies inside the horizon and is regular. The geometry describes a black hole whose interior continues smoothly through $r=0$ into a separate asymptotic region, rather than terminating at a singularity.

    \item \textbf{Extremal Black-Bounce ($a = 2m$):} The coordinate horizon location shrinks to $r_+ = 0$, so that the horizon coincides with the bounce surface. This surface is regular: the Kretschmann scalar evaluates to $K(0) = 12m^2/a^6 = 3/(4m^4)$, which is finite. The surface gravity vanishes ($\kappa = 0$) and the geometry represents a one-way wormhole with an extremal null throat .
    
    \item \textbf{Traversable Wormhole ($a > 2m$):} The condition $\sqrt{r^2 + a^2} = 2m$ has no real solution since $a > 2m$ implies $\mathcal{R}(r) \geq a > 2m$ for all $r$. No horizon exists and $f(r) > 0$ everywhere. The surface $r=0$ is a regular, timelike, two-way traversable throat with minimal areal radius $\mathcal{R}(0) = a$ .
    
    \item \textbf{Schwarzschild Limit ($a \to 0$):} The coordinate horizon location approaches $r_+ \to 2m$, recovering the standard Schwarzschild event horizon. In this limit alone, the bounce surface degenerates, $\mathcal{R}(0) = a \to 0$ and the curvature invariants diverge ($K(0) = 12m^2/a^6 \to \infty$), recovering the Schwarzschild singularity. This singular case is a degenerate boundary of the parameter space; the entire analysis in this work concerns $a > 0$.

\end{itemize}

\begin{figure}[h!]
    \centering
    \includegraphics[width=0.48\textwidth]{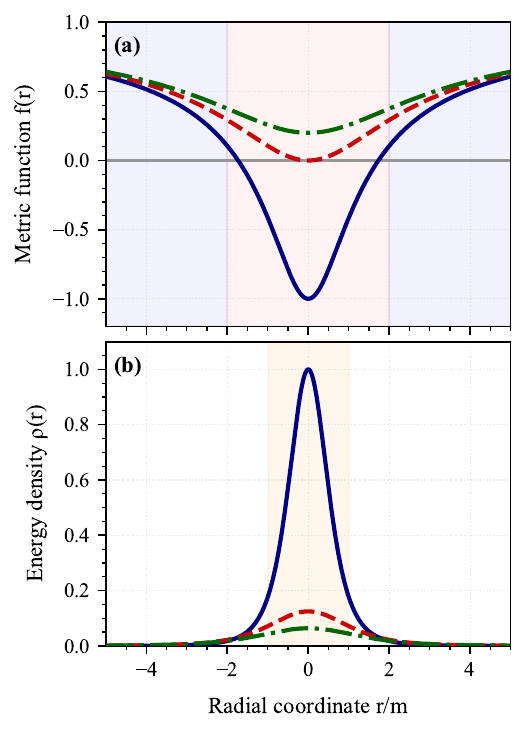}
    \caption{Metric structure and energy density for the Simpson--Visser spacetime with mass $m = 1$. \textbf{(a)} The metric function $f(r)$ illustrates the transition from a regular black hole ($a=1.0$) to an extremal case ($a=2.0$) and finally to a traversable wormhole ($a=2.5$). \textbf{(b)} The corresponding energy density $\rho(r)$ shows regular, localized matter supporting the geometry. Horizon structure and energy profiles are directly controlled by the parameter $a$. In both panels, the lines correspond to different values of the regularization parameter $a$: the solid blue line represents the regular black hole case ($a=1.0$), the dashed red line represents the extremal black-bounce ($a=2.0$), and the dash-dot green line represents the traversable wormhole ($a=2.5$).}
    \label{fig:structure}
\end{figure}

\subsection{Stress-Energy Tensor and Energy Conditions}

The geometry \eqref{eq:metric} is a solution to the Einstein field equations, $G_{\mu\nu} = 8\pi T_{\mu\nu}$, for a specific physical source. The physical nature of the supporting matter is best analyzed by the energy conditions \cite{Balart:2014jia}. The Null Energy Condition (NEC), which states $T_{\mu\nu}k^\mu k^\nu \ge 0$ for any null vector $k^\mu$, is the most fundamental of these. For a spherically symmetric fluid, this requires $\rho+p_\parallel \ge 0$ and $\rho+p_\perp \ge 0$. Using the Einstein tensor from the source paper \cite{Simpson:2018tsi}, we find:
\begin{equation}
\rho + p_\parallel = - \frac{a^2 ( \sqrt{r^2+a^2}-2m)}{4\pi (r^2+a^2)^{5/2}}.
\label{eq:nec_radial}
\end{equation}

This term holds for all values of $r$, that is \textit{inside} and \textit{outside} the horizon, and is negative everywhere except on any horizon which might be present, indicating a clear violation of the NEC. This violation of NEC is sufficient to conclude that the weak, strong, and dominant energy conditions shall also be violated. This violation is a feature and  requirement that singularity resolution within classical general relativity is impossible without the presence of such ``exotic'' matter.\cite{Morris:1988tu,Visser:1989kh,Visser:1995cc}\\

\subsection{Semiclassical Thermodynamics}

We now establish the thermodynamic properties of the black hole in the semiclassical approximation ($0 < a \le 2m$). The Hawking temperature, $T_H = \kappa/(2\pi)$, is derived from the surface gravity $\kappa$ at the outer event horizon, $r_+ = \sqrt{4m^2-a^2}$. The surface gravity is correctly calculated as:
\begin{equation}
\kappa = \frac{1}{2} \left. \frac{d f(r)}{dr} \right|_{r=r_+} = \frac{1}{2} \left[ \frac{2mr}{(r^2+a^2)^{3/2}} \right]_{r=r_+} = \frac{m r_+}{(r_+^2+a^2)^{3/2}}.
\label{eq:kappa_deriv}
\end{equation}
Substituting the horizon condition $\sqrt{r_+^2+a^2} = 2m$, we obtain:
\begin{equation}
\kappa = \frac{m \sqrt{4m^2-a^2}}{(2m)^3} = \frac{\sqrt{4m^2-a^2}}{8m^2}.
\label{eq:kappa_final}
\end{equation}
The Hawking temperature is therefore
\begin{equation}
T_H = \frac{\kappa}{2\pi} = \frac{\sqrt{4m^2-a^2}}{16\pi m^2}.
\label{eq:hawking_temp}
\end{equation}

This expression exhibits the correct physical limits. For $a \to 0$, it recovers the standard Schwarzschild temperature, $T_H \to 1/(8\pi m)$. For the extremal case $a \to 2m$, the temperature vanishes, $T_H \to 0$, as expected for an extremal black hole.

The Bekenstein-Hawking entropy is obtained from the area of the outer event horizon. Following the standard procedure \cite{Ali:2025FdP}, the horizon area is computed from the angular part of the metric as
\[
A_+ = \int_0^{2\pi} \int_0^{\pi} \sqrt{g_{\theta\theta}\, g_{\phi\phi}}\; d\theta\, d\phi \,\bigg|_{r=r_+}.
\]
For the Simpson--Visser metric \eqref{eq:metric}, the angular metric components at the horizon are $g_{\theta\theta} = r_+^2 + a^2$ and $g_{\phi\phi} = (r_+^2 + a^2)\sin^2\theta$. Evaluating the integral yields
\[
A_+ = 4\pi (r_+^2 + a^2).
\]
Using the horizon condition $r_+^2 + a^2 = (2m)^2$, this simplifies to
\[
A_+ = 16\pi m^2.
\]
The Bekenstein-Hawking entropy then follows as
\begin{equation}
S_{BH} = \frac{A_+}{4} = 4\pi m^2.
\label{eq:bh_entropy}
\end{equation}
The $a$-independence of this result deserves careful examination. It is a direct consequence of the horizon condition $\mathcal{R}(r_+) = \sqrt{r_+^2 + a^2} = 2m$, which fixes the areal radius at the horizon to $2m$ regardless of the value of $a$. Consequently, the Bekenstein-Hawking entropy is degenerate across the entire family of regular black holes of the same mass, a near-Schwarzschild black hole ($a \ll 2m$) and an extremal black-bounce ($a = 2m$) share identical semiclassical entropy. This degeneracy is, however, an artifact of applying the area law directly at the horizon. As we demonstrate in Section~\ref{sec:quantum_entropy}, when one correctly accounts for the $a$-dependent energy-temperature relation through thermodynamic integration, the resulting entropy explicitly depends on $a$ even at leading order, breaking this degeneracy and revealing that the entropy is sensitive to the regular core structure. These foundational quantities, $T_H(m,a)$ and $S_{BH}(m)$, provide the essential starting point for this investigation.

\section{Thermodynamic Stability and Phase Transition}
\label{sec:stability}

The thermodynamic response of a black hole, encapsulated by its heat capacity, provides deep insights into its stability and its ability to achieve equilibrium with a thermal environment. In a canonical ensemble, a positive heat capacity is a necessary condition for thermodynamic stability, whereas a negative capacity signals an instability. \cite{Davies:1978}. The Schwarzschild black hole, with its negative heat capacity, becomes hotter as it radiates energy. Black hole phase transitions, including Van der Waals–like behavior and critical phenomena, have been widely analyzed in both singular and regular spacetimes \cite{Witten:1998zw,Kastor:2009wy,Ali:2019rjn,Ruppeiner:2018pgn,Kubiznak:2012wp}. In this section, we investigate how the regularization parameter $a$ of the Simpson--Visser model fundamentally alters this behavior, revealing that this regular spacetime exhibits a critical instability, marked by a Davies point where the system undergoes a thermodynamic stability transition."

\subsection{Heat Capacity of the Regular Black Hole}

The heat capacity, for a system with zero charge and angular momentum, is defined as the response of the system's energy to a change in its temperature. Identifying the black hole's internal energy with its ADM mass, $E=m$, the heat capacity is given by:
\begin{equation}
    C = \left(\frac{\partial T_H}{\partial m}\right)^{-1}.
    \label{eq:heat_capacity_def}
\end{equation}
To compute this, we must first evaluate the derivative of the Hawking temperature, Eq.~\eqref{eq:hawking_temp}, with respect to the mass $m$:
\begin{equation}
    \frac{\partial T_H}{\partial m} = \frac{\partial}{\partial m} \left( \frac{\sqrt{4m^2 - a^2}}{16\pi m^2} \right).
\end{equation}
A straightforward application of the quotient rule yields:
\begin{align}
    \frac{\partial T_H}{\partial m} &= \frac{1}{16\pi} \left[ \frac{m^2 \left( \frac{4m}{\sqrt{4m^2-a^2}} \right) - (\sqrt{4m^2-a^2})(2m)}{m^4} \right] \nonumber \\
    &= \frac{2a^2 - 4m^2}{16\pi m^3 \sqrt{4m^2 - a^2}}.
\end{align}
Taking the reciprocal to find the heat capacity, we arrive at our central result for this section:
\begin{equation}
    C(m,a) = \frac{16\pi m^3 \sqrt{4m^2 - a^2}}{2a^2 - 4m^2} = -\frac{8\pi m^3 \sqrt{4m^2 - a^2}}{2m^2 - a^2}.
    \label{eq:heat_capacity_final}
\end{equation}
This expression reveals a thermodynamic structure. Unlike the strictly negative heat capacity of the Schwarzschild black hole, recovered by taking the limit $a \to 0$, which yields $C \to -8\pi m^2$, the sign and behavior of $C(m,a)$ for the Simpson--Visser black hole are critically dependent on the value of the regularization parameter $a$. Works by Bargueño \textit{et.al} reinterprets such change of sign of the specific heat which takes place through an infinite discontinuity. Such thermodynamic phase transitions occurs at so called Davies points \cite{Bargueno:2024mys}. 

\subsection{Stability Analysis and Davies Type Phase Transition}

The most striking feature of the heat capacity in Eq.~\eqref{eq:heat_capacity_final} is the pole originating from its denominator, $2m^2 - a^2$. The heat capacity diverges when this term vanishes, signaling a phase transition in the system. This divergence occurs at a critical value of the regularization parameter, $a_{\text{crit}}$, defined by:
\begin{equation}
    2m^2 - a_{\text{crit}}^2 = 0 \quad \implies \quad a_{\text{crit}} = \sqrt{2} m.
    \label{eq:critical_point}
\end{equation}
This critical point lies within the allowed range for a regular black hole, $0 < a_{\text{crit}} < 2m$. The divergence of the heat capacity is the characteristic of a \textbf{phase transition} \cite{Davies:1977,Sokolowski:1980uva}. This transition separates two distinct thermodynamic phases, whose stability is dictated by the sign of the heat capacity.

\subsubsection{The Unstable Phase ($0 < a < \sqrt{2}m$)}
In this regime, the denominator $2m^2 - a^2$ in Eq.~\eqref{eq:heat_capacity_final} is positive. Since the numerator is manifestly negative for any non-extremal black hole ($a < 2m$), the heat capacity $C$ is negative. Black holes in this phase are thermodynamically unstable. They behave qualitatively like the Schwarzschild black hole, unable to coexist in stable equilibrium with a thermal reservoir and subject to runaway evaporation, becoming progressively hotter as they lose mass.

\subsubsection{The Stable Phase ($\sqrt{2}m < a < 2m$)}
When the regularization parameter exceeds the critical value, the denominator $2m^2 - a^2$ becomes negative. The heat capacity becomes positive ($C>0$). Consequently, black holes in this phase are thermodynamically stable. They can achieve stable thermal equilibrium with a surrounding heat bath, absorbing and emitting radiation to maintain a constant temperature. This behavior is a departure from the classical black hole picture and is a direct physical consequence of the modified spacetime geometry near the regular core.

\begin{figure}[ht]
    \centering
    \includegraphics[width=0.5\textwidth]{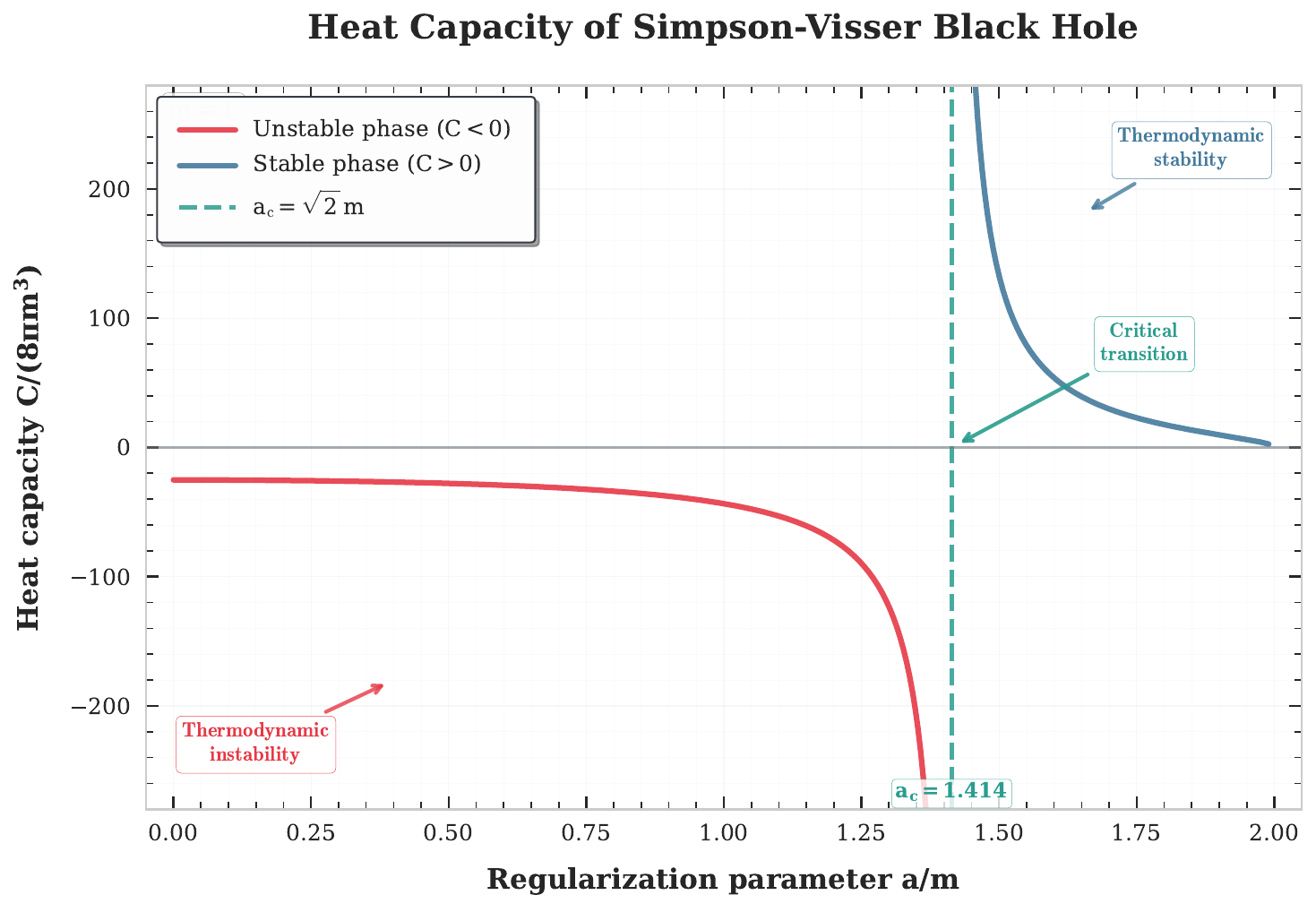} 
    \caption{Normalized heat capacity as a function of the dimensionless regularization parameter $(a/m)$. The plot reveals the existence of two distinct thermodynamic phases. For $a < \sqrt{2}m$, the heat capacity is negative (blue shaded region), corresponding to an unstable phase analogous to the Schwarzschild black hole. For $\sqrt{2}m < a < 2m$, the heat capacity is positive (red shaded region), indicating a thermodynamically stable phase. The vertical dashed line marks the critical point where the heat capacity diverges, signifying a phase transition between the stable and unstable phases. }
    \label{fig:heat_capacity}
\end{figure}

The transition at $a=a_{\text{crit}}$ therefore represents a shift in the black hole's physical nature. The regularization parameter $a$ is not merely a geometric cutoff but acts as a parameter that drives the system across a critical point, transforming it from an unstable, perpetually evaporating object into a stable one. This thermodynamic behavior is illustrated in Fig:(\ref{fig:heat_capacity})

The free energy is defined as
\begin{equation}
F = E - TS,
\end{equation}
where the internal energy is identified with the ADM mass, \( E = m \). Using the black hole entropy \( S = 4\pi m^2 \) and the Hawking temperature
\[
T(a,m) = \frac{\sqrt{4m^2 - a^2}}{16\pi m^2},
\]
we obtain the free energy as
\begin{align}
F(a, m) &= m - T(a,m) S(m) \nonumber \\
       &= m - \left( \frac{\sqrt{4m^2 - a^2}}{16\pi m^2} \right) (4\pi m^2) \nonumber \\
       &= m - \frac{\sqrt{4m^2 - a^2}}{4}.
       \label{freeenrgy}
\end{align}
This expression reveals that the free energy monotonically decreases with increasing regularization parameter \( a \), vanishing in the extremal limit \( a \to 2m \). The result is consistent with the decreasing temperature and increasing stability near the wormhole threshold.
\begin{figure}[hb]
    \centering
    \includegraphics[width=0.48\textwidth]{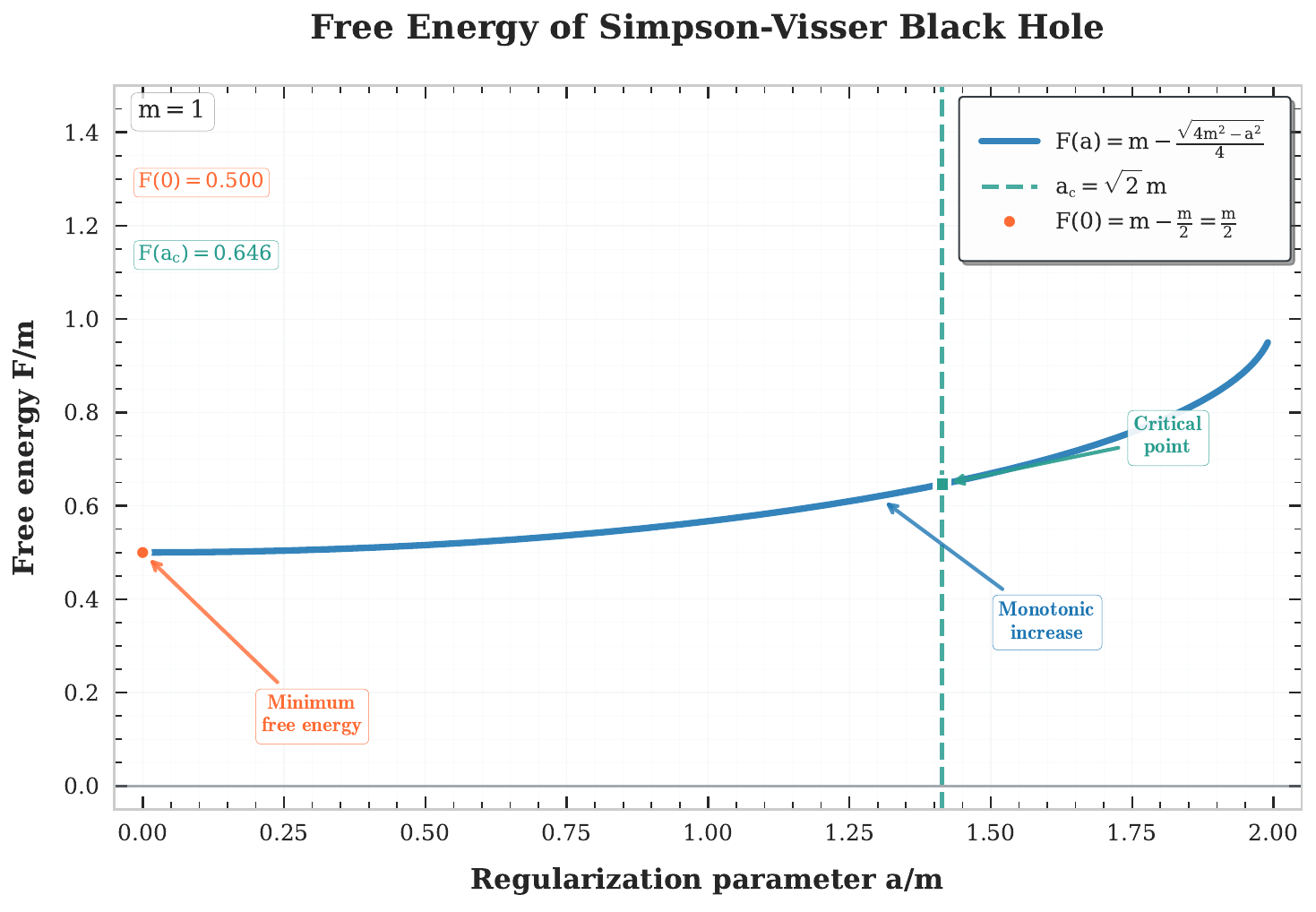}
    \caption{The free energy indicates the thermodynamically favored state of the system. The plot reveals that the free energy is an increasing function of $a$, achieving its global minimum of $F=m/2$ at $a=0$. This implies that the singular Schwarzschild configuration is the thermodynamically preferred state among all regular black hole solutions of the same mass. The physical thermodynamic evolution is instead governed by mass loss due to Hawking evaporation at fixed $a$. As the mass $m$ decreases, the system crosses a critical point at $m=a/\sqrt{2}$, beyond which the heat capacity becomes positive. This signals a change in the thermal response. The vertical dashed line marks the critical point $a_c = \sqrt{2}m$ of the phase transition for reference.}
    \label{fig:free_energy}
\end{figure}

\section{Quantum Entropy Corrections from the Tunneling Formalism}
\label{sec:quantum_entropy}

The thermodynamic framework established in the preceding sections, while revealing a rich phase structure, is fundamentally semiclassical. The Bekenstein-Hawking entropy, $S_{BH} = A/4G\hbar$, represents the leading-order contribution in a low-energy effective field theory description of gravity. It is widely anticipated that a complete theory of quantum gravity will introduce subleading corrections to this area law. These corrections are expected to encode the profound information about the quantum nature of spacetime, the statistical origin of entropy, and the possible resolution of the information paradox.

In this section, we advance beyond the semiclassical limit to compute these quantum corrections for the Simpson--Visser regular black hole. To achieve this, we employ the tunneling formalism, that models Hawking radiation as a quantum tunneling process across the event horizon. Our primary objective is to derive an expression for the entropy that incorporates quantum back-reaction effects and to analyze how the singularity-resolving parameter $a$ influences the quantum-statistical properties of the black hole.

\subsection{The Hamilton-Jacobi Method and Quantum Corrections}

The conceptual foundation of our approach is the modeling of Hawking radiation as particle tunneling, a picture pioneered by Parikh and Wilczek \cite{Parikh:1999mf,Parikh:2004ih}. In the purely semiclassical analysis, a particle is treated as moving in a fixed, unperturbed spacetime background. The probability for this particle to tunnel across the classically forbidden region of the event horizon is given by the WKB approximation, $\Gamma \sim \exp(-2 \text{Im}[I_0]/\hbar)$, where $I_0$ is the leading-order classical action for the particle's trajectory. This method successfully reproduces the standard Hawking temperature, but it doesn't take much account into the back-reaction of the emitted particle on the spacetime geometry itself. 

To construct a more precise picture, these back-reaction effects must be systematically incorporated. A method for achieving this is the Hamilton-Jacobi formalism extended to include higher-order quantum corrections, as developed by Banerjee and Majhi \cite{Banerjee:2008cf,Majhi:2009pr}. The central idea is to expand the particle's action, $I$, as a power series in $\hbar$:
\begin{equation}
    I = I_0 + \hbar I_1 + \hbar^2 I_2 + \dots
\end{equation}

Physically, this expansion corresponds to moving beyond the test-particle approximation. The term $I_0$ represents the action on a fixed background, while the higher-order terms $I_n$ account for the dynamic response of the geometry and self-interaction effects.

Substituting this expansion into the WKB tunneling probability $\Gamma \sim \exp(-2\,\mathrm{Im}[I]/\hbar)$ gives
\[
\Gamma \sim \exp\!\left(-\frac{2\,\mathrm{Im}[I_0]}{\hbar} - 2\,\mathrm{Im}[I_1] - 2\hbar\,\mathrm{Im}[I_2] - \cdots\right).
\]
The leading term $\mathrm{Im}[I_0]$ reproduces the standard Boltzmann factor $\Gamma \sim e^{-E/T_H}$ and yields the semiclassical Hawking temperature. The higher-order terms $I_1, I_2, \ldots$ encode the back-reaction of the emitted particle on the spacetime geometry and self-gravitational effects: each radiated quantum carries away energy, causing the black hole mass and horizon to shrink, and these terms systematically account for this dynamic response. Identifying the full tunneling probability with the Boltzmann factor $\Gamma \sim e^{-E/T_{\text{corr}}}$ and comparing order by order in $\hbar$, one obtains a quantum-corrected temperature \cite{Banerjee:2008cf,Majhi:2009pr}:
\begin{equation}
    T_{\text{corr}} = T_H \left( 1 + \sum_{i=1}^\infty \beta_i \frac{\hbar^i}{m^{2i}} \right)^{-1}.
\end{equation}
The dimensionless coefficients $\beta_i$ are phenomenological parameters that encapsulate the specific details of the underlying quantum gravity theory. They are expected to be of order unity. For our analysis, we will consider corrections up to the $\mathcal{O}(\hbar^2)$ term, as this is sufficient to capture the leading logarithmic and inverse-mass corrections to the entropy.

Quantum corrections to black hole thermodynamics have also been extensively studied in the literature through an alternative route based on the generalized uncertainty principle (GUP). This approach has been applied to a variety of black hole spacetimes, including rotating regular Hayward black holes \cite{Ali:2022PDU}, Kiselev-like AdS black holes in $f(R,\,T)$ gravity \cite{Ali:2025pdu}, and charged black holes in $f(R)$ gravity \cite{Ali:2025FdP}. The resulting corrections to the Hawking temperature in those analyses depend on the GUP parameter and the properties of the emitted particle.

In the present work, we follow a complementary approach developed by Banerjee and Majhi \cite{Banerjee:2008cf,Majhi:2009pr}, where the quantum corrections arise from expanding the particle's action as a power series in $\hbar$ within the Hamilton-Jacobi formalism. In this framework, the correction coefficients $\beta_i$ encode back-reaction and self-gravitational effects of the emitting process on the background geometry, and the resulting corrected entropy is obtained through thermodynamic integration of the first law, $dS = dm/T_{\text{corr}}$. Both approaches yield logarithmic corrections to the entropy at leading order, consistent with results from loop quantum gravity and the Cardy formula \cite{Kaul:2000kf,Carlip:2000nv}.

With the quantum-corrected temperature in hand, we can now derive the corresponding corrected entropy, $S$, by integrating the first law of black hole thermodynamics, $dS = dM/T$. Identifying the black hole energy $M$ with its mass $m$, we have:
\begin{equation}
    S = \int \frac{dm}{T_{\text{corr}}} = \int \frac{1}{T_H} \left( 1 + \frac{\beta_1 \hbar}{m^2} + \frac{\beta_2 \hbar^2}{m^4} + \dots \right) dm.
\end{equation}
This integral forms the basis of our calculation.

\subsection{The Schwarzschild Limit}

Before proceeding to the full calculation for the Simpson--Visser spacetime, we first see this methodology on the well-understood Schwarzschild case. This serves as a crucial benchmark and a check of consistency. We set the regularization parameter $a=0$ in the integrand. The Hawking temperature for a Schwarzschild black hole is $T_H = \hbar/(8\pi k_B m)$. The entropy integral thus becomes:
\begin{equation}
S_{\text{Schw}}(m) = \int \frac{8\pi k_B m}{\hbar} \left( 1 + \frac{\beta_1 \hbar}{m^2} + \frac{\beta_2 \hbar^2}{m^4} \right) dm.
\end{equation}
We can evaluate this term by term:
\begin{equation}
S_{\text{Schw}}(m) = \frac{8\pi k_B}{\hbar} \int \left( m + \frac{\beta_1 \hbar}{m} + \frac{\beta_2 \hbar^2}{m^3} \right) dm.
\end{equation}
The integration is straightforward:
\begin{equation}
S_{\text{Schw}}(m) = \frac{8\pi k_B}{\hbar} \left( \frac{m^2}{2} + \beta_1 \hbar \ln(m) - \frac{\beta_2 \hbar^2}{2m^2} \right) + S_{\text{const}}.
\end{equation}
Re-arranging the terms and expressing the leading term using the Bekenstein-Hawking entropy $S_{BH} = 4\pi k_B m^2/\hbar$, we find:
\begin{equation}
S_{\text{Schw}}(m) = S_{BH} + 8\pi k_B \beta_1 \ln(m) - \frac{4\pi k_B \beta_2 \hbar}{m^2} + S_{\text{const}}.
\end{equation}
This result, featuring the area law, a leading logarithmic correction, and an inverse-mass correction, matches the established results in the literature for quantum-corrected Schwarzschild entropy \cite{Banerjee:2008cf, Kaul:2000kf,Bianchi:2012ui,Sen:2012dw,Carlip:2000nv}. 
We now apply it to the regular, non-singular spacetime.

\subsection{Quantum-Corrected Entropy of the Simpson--Visser Black Hole}

We now return to the full expression for the Simpson--Visser case, with $a \neq 0$. Throughout this analysis we restrict to the regular black hole regime $0<a<2m$, for which an event horizon exists and the Parikh–Wilczek tunneling method applies. The integral to be solved is:
\begin{equation}
S(m,a) = \frac{16\pi k_B}{\hbar} \int \left( 1 + \frac{\beta_1 \hbar}{m^2} + \frac{\beta_2 \hbar^2}{m^4} \right) \frac{m^2}{\sqrt{4m^2 - a^2}} dm.
\end{equation}
We can separate it into three distinct integrals, $S = S_0 + S_1 + S_2$.

The first integral, $S_0$, corresponds to the modified semiclassical contribution:
\begin{equation}
S_0 = \frac{16\pi k_B}{\hbar} \left[ \frac{m}{8}\sqrt{4m^2 - a^2} + \frac{a^2}{16}\ln\left(2m + \sqrt{4m^2 - a^2}\right) \right].
\end{equation}
The second integral, $S_1$, provides the leading-order quantum correction, which we expect to be logarithmic:
\begin{equation}
S_1 = 8\pi k_B \beta_1 \ln\left(2m + \sqrt{4m^2 - a^2}\right).
\end{equation}
The third integral, $S_2$, gives the next-to-leading order correction:
\begin{equation}
S_2 = - \frac{16\pi k_B \beta_2 \hbar \sqrt{4m^2 - a^2}}{a^2 m}.
\end{equation}
Combining these three results, we arrive at the final expression for the quantum-corrected entropy of the Simpson--Visser regular black hole:
\begin{widetext}
\begin{equation}
\label{eq:final_corrected_entropy_full}
\begin{split}
S(m,a) = \frac{2\pi k_B}{\hbar} \Bigg( 
    m \sqrt{4m^2 - a^2} 
    + \frac{a^2}{2} \ln\left( \frac{2m + \sqrt{4m^2 - a^2}}{a} \right) 
\Bigg)
+ 8\pi k_B \beta_1 \ln\left( \frac{2m + \sqrt{4m^2 - a^2}}{a} \right) \\
- \frac{16\pi k_B \beta_2 \hbar \sqrt{4m^2 - a^2}}{a^2 m} 
+ S_{\text{const}}.
\end{split}
\end{equation}
\end{widetext}

\begin{figure*}[t!]
    \centering
    \includegraphics[width=0.7\textwidth,height=0.5\textwidth]{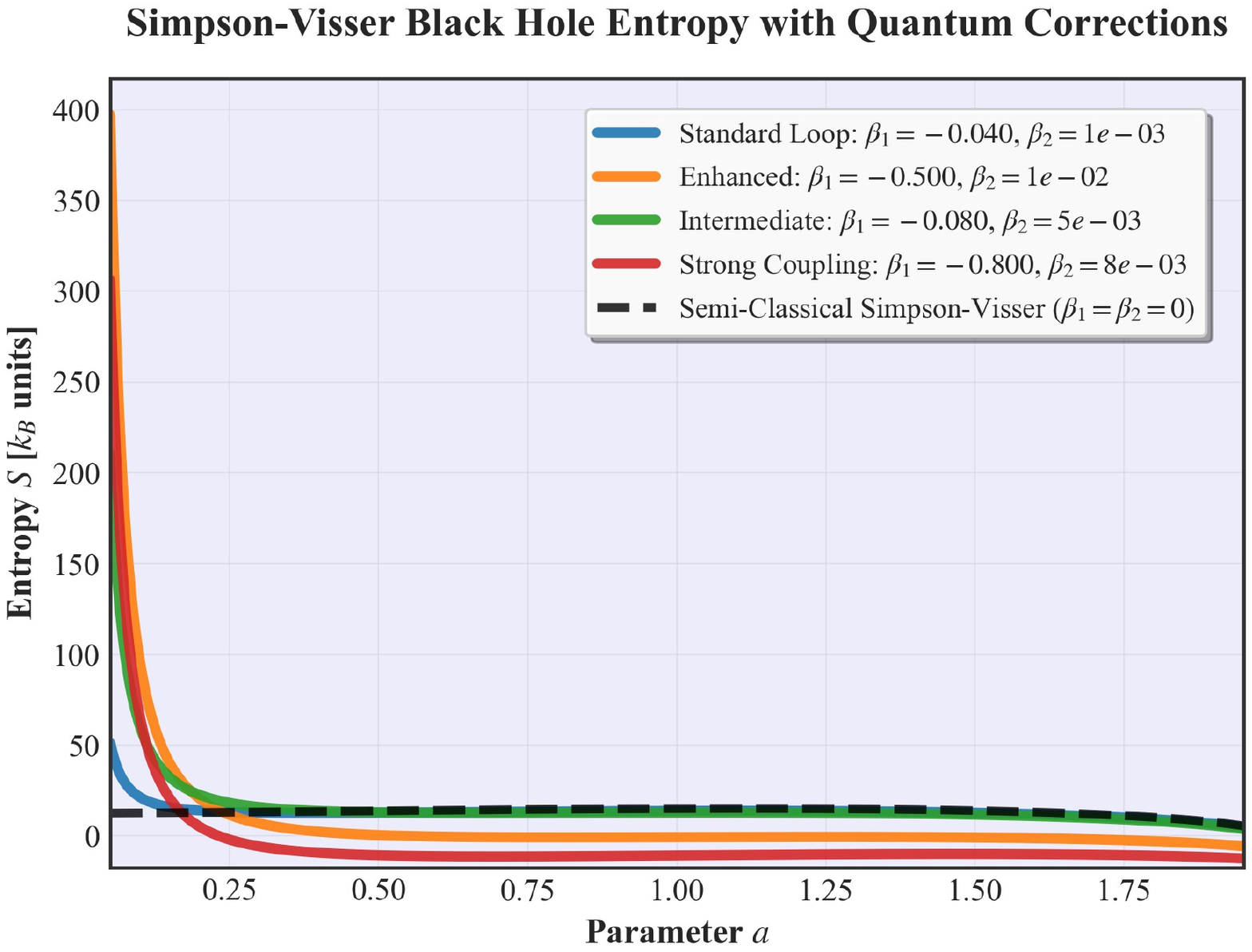}
    \caption{The quantum-corrected entropy $S$ of the Simpson--Visser black hole as a function of the regularization parameter $a$, calculated for a fixed mass $m=1$. The different colored lines represent distinct phenomenological scenarios for the quantum correction coefficients $(\beta_1, \beta_2)$, as specified. The dashed black line indicates the Bekenstein-Hawking area-law entropy $S_{BH} = 4\pi m^2$, which coincides with the Schwarzschild value and is independent of $a$. The plot demonstrates that quantum corrections can significantly alter the entropy, particularly for small $a$ where the geometry is nearly singular. For certain choices of parameters (e.g., strong coupling), the total entropy can become negative, signaling the need for non-perturbative effects in that regime. All corrected entropies converge towards the semiclassical result for large $a$ near the extremal limit. Standard coupling is for the coupling constant $\beta =\frac{-1}{8\pi}$. The enhanced, intermediate and strong coupling are taken arbitrarily to demonstrate the effects of coupling constant on the entropy.}
    \label{fig:main_entropy}
\end{figure*}

\begin{figure*}[t!]
    \centering
    \includegraphics[width=0.8\textwidth]{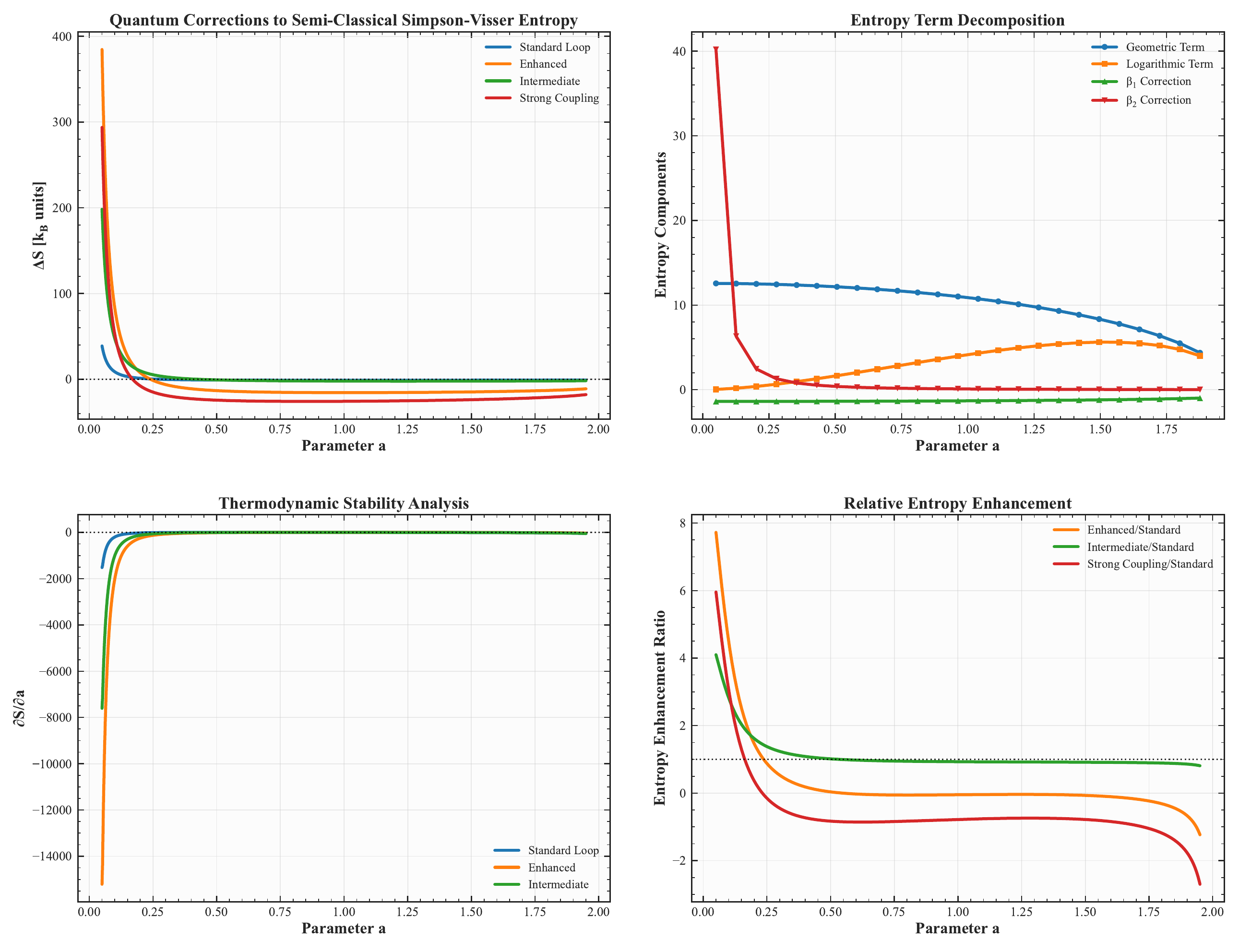}
    \caption{A detailed decomposition and analysis of the quantum-corrected entropy for various phenomenological scenarios. 
    \textbf{(a)} The deviation of the corrected entropy from the semiclassical value, $\Delta S = S - S_{BH}$, isolating the net effect of the quantum corrections. 
    \textbf{(b)} A decomposition of the total entropy into its constituent parts: the modified semiclassical geometric term ($S_0$), the logarithmic correction term ($S_1 \propto \beta_1$), and the higher-order inverse-mass term ($S_2 \propto \beta_2$). This panel reveals the dominant contribution of the higher-order term at small $a$. 
    \textbf{(c)} The thermodynamic stability derivative, $\partial S / \partial a$, indicating how the total entropy changes with the regularization parameter. 
    \textbf{(d)} The relative entropy enhancement ratio, showing the factor by which quantum corrections enhance or suppress the total entropy relative to the semiclassical value.}
    \label{fig:analysis_subplots}
\end{figure*}

\subsection{Physical Analysis and Interpretation of the Corrected Entropy}
This final expression is rich with physical meaning and deserves a detailed dissection. It reveals how the act of resolving the singularity permeates every level of the thermodynamic description.\\

\paragraph{The Modified Semiclassical Term.} The first term of the result is the leading-order term, $S_0$. It is crucial to note that this is \textit{not} the Bekenstein-Hawking area law we obtained earlier from equation (\ref{eq:bh_entropy}), $S_{BH} = 4\pi k_B m^2/\hbar$. Instead, it is a more complex function that explicitly depends on the regularization parameter $a$. This is a profound point: the thermodynamic integration, which correctly accounts for the energy-temperature relation across all mass scales, reveals that the entropy ``knows'' about the regular core structure even at the leading order. \\

\paragraph{The Logarithmic Correction.} The second term, proportional to $\beta_1$, represents the leading-order quantum correction. Its logarithmic form is a feature predicted by a wide variety of approaches to quantum gravity, including loop quantum gravity, string theory, and general arguments based on the Cardy formula, earlier calculation shows $\beta_1=-\frac{1}{8\pi}$ \cite{Kaul:2000kf, Carlip:2000nv,Fursaev:1994te,Mann:1997hm,Das:2001ic,Ghosh:2004wq}. This term is believed to count the quantum states responsible for the entropy. Our result provides the specific form of this correction for a regular black hole, showing how the argument of the logarithm is determined by the interplay between the mass $m$ and the regularization scale $a$.\\

\paragraph{The Higher-Order Correction and the Role of \texorpdfstring{$a$}{a}.} The third term, proportional to $\beta_2$, is the next subleading correction. It has a highly non-trivial dependence on $a$, scaling as $1/a^2$. This implies that for very small values of $a$ (i.e., for black holes that are very close to the singular Schwarzschild limit), this correction term can become very large. It highlights the role of $a$ as a physical regulator not just for the geometry, but for the convergence of the quantum thermodynamic description itself.\\

\paragraph{Implications for the Black Hole Final State.} The complete expression for entropy, combined with our finding of a stable thermodynamic phase in Section III, provides a picture of the end-stage of black hole evaporation. As the black hole evaporates and its mass $m$ decreases, it eventually approaches the stable regime ($m < a/\sqrt{2}$). The evaporation would then as a result slows down and eventually cease, leaving an extremal remnant. As the mass $m$ decreases toward the extremal condition $m = a/2$, the black hole approaches a zero-entropy state. Consequently, the entropy derived via the tunneling formalism vanishes:

\begin{equation}
    \lim_{m \to a/2} S(m, a) \to 0.
\end{equation}

\section{Discussion and Conclusion}
\label{sec:conclusion}

The breakdown of physical laws at the curvature singularity has long been a primary motive for seeking a quantum theory of gravity. This work confronted this foundational problem by undertaking a comprehensive thermodynamic and quantum analysis of the Simpson--Visser spacetime. Our investigation was driven by the physical idea that the resolution of the singularity should not be merely a geometric convenience but a profound physical event that must imprint detectable, non-trivial signatures upon the black hole's thermodynamic and quantum properties. Our findings confirms this in a series of interconnected results that describes a new, self-consistent picture of the black hole life cycle.

Our first major result was the discovery of a Davies Point transition at a critical regularization scale, $a_{\text{crit}} = \sqrt{2}m$. This transition separates an unstable, Schwarzschild-like thermodynamic phase ($a < a_{\text{crit}}$) from a thermodynamic stable phase ($a > a_{\text{crit}}$). This result implies that the regularization of the singularity fundamentally restructures the black hole’s thermodynamic landscape, permitting it to achieve stable thermal equilibrium with a surrounding environment, heat bath, a possibility forbidden for its singular counterpart. However, our analysis of the free energy added a crucial layer of physical nuance. We found that the free energy is minimized at $a=0$, indicating that the singular Schwarzschild state is the preferred thermodynamic configuration. However, this minimum is physically inaccessible if $a$ represents a fundamental length scale of the spacetime. In this scenario, the black hole cannot decay to the singular state. Instead, its thermodynamic evolution is driven by mass loss due to evaporation. As the mass m decreases for a fixed a, the system is driven towards the thermodynamically stable regime $(a > \sqrt{2}m)$, where the heat capacity becomes positive.

The second result of our investigation was the derivation of the quantum-corrected entropy via the Hamilton-Jacobi tunneling formalism. Here, our analysis yielded several insights. Perhaps the most subtle but interesting outcome was that the leading-order entropy, when derived through employing Hamilton-Jacobi approach, explicitly depends on the regularization parameter $a$. These results show that the thermodynamics is sensitive to the presence of the regular core from the very beginning, contradicting the expectation that the core only modifies subleading corrections. Such dependence is concealed if one applies the Bekenstein–Hawking area law directly at the horizon $r_+$. Furthermore, the higher-order correction term scales as $1/a^2$. Its divergence for small $a$ signals that the perturbative quantum expansion becomes incomplete precisely when the regular core is very small and the spacetime approaches its singular limit. This suggests that the regularization parameter $a$ functions not only as a geometric regulator but also as a regulator for the validity of the effective field theory description of the quantum entropy.

Synthesizing these two threads, the thermodynamic phase structure and the corrected entropy, provides a self-consistent narrative for the end-stage of black hole evaporation. A large black hole initially in the unstable phase ($a \ll m$) will radiate and shrink. As its mass decreases, it eventually crosses the critical point into the stable phase, where its evaporation would dramatically slow. The process ceases when the black hole reaches the zero-temperature extremal state at $a=2m$, leaving a stable remnant. Our work provides the statistical-mechanical description of this remnant. Its entropy, $S_{\text{remnant}} \propto \beta_1 \ln(a)$, is a non-zero, logarithmic quantity determined by the quantum gravity scale $a$ and the leading quantum correction coefficient $\beta_1$. The physical relevance of such stable endpoints has been emphasized in earlier studies of regular black holes, particularly in the context of early-universe physics and phase transitions. In a series of works, Khlopov and collaborators have shown that regular black hole remnants, especially those with de Sitter–like interiors, can arise from primordial processes and may have important cosmological and phenomenological relevance \cite{Khlopov:2008qy,Dymnikova:2015yma}. These studies provide physical motivation for treating regular black hole remnants as meaningful end states of evaporation. In this context, our results offer a complementary thermodynamic perspective. We show that in the Simpson–Visser black-bounce geometry the regularization scale itself acts as a control parameter governing thermodynamic stability, driving a phase transition that separates unstable and stable regimes. When quantum back-reaction effects are incorporated through the tunneling formalism, this framework provides a thermodynamic characterization of the extremal endpoint, supporting the view that regular black hole remnants can arise as dynamically selected, stable configurations.

This work opens several important avenues for future research. An essential extension in this can be the inclusion of rotation, in order to explore the interplay between angular momentum, phase transitions, and the structure of the remnant. Equally interesting is the dynamical analysis of gravitational collapse into Simpson–Visser–like geometries, to determine whether the stable configurations identified here can be observed. More broadly, by adopting singularity resolution as a guiding physical principle, we have uncovered a rich thermodynamic structure characterized by phase transitions and quantum corrections. In doing so, we establish regular black holes as thermodynamically active systems, providing a framework for a self-consistent, singularity-free description of the black hole life cycle.

\section*{Acknowledgements}

The authors thank Rajes Ghosh for his valuable comments and helpful suggestions that contributed to improving this work.

\bibliographystyle{apsrev4-2}
\bibliography{references}

\end{document}